\documentstyle[epsfig]{aipproc}

\begin{document}


\newcommand{\EEG}{\rm e^+ e^-\rightarrow \gamma\gamma}
\newcommand{\EEGG}{\rm e^+ e^-\rightarrow \gamma\gamma(\gamma)}
\newcommand{\EEGGG}{\rm e^+ e^-\rightarrow \gamma\gamma\gamma}
\newcommand{\EEEEG}{\rm e^+ e^-\rightarrow e^+e^-(\gamma)}
\newcommand{\LAMP}{ \Lambda_{+}}
\newcommand{\LAMM}{ \Lambda_{-}}
\newcommand{\LAMS}{ \Lambda_{6}}
\newcommand{\LAMPP}{ \Lambda_{++}}
\newcommand{\LAMMM}{ \Lambda_{--}}
\newcommand{\DSDW}{ \sigma(\theta)}
\newcommand{\MESTAR}{ m_{{\rm e^{\ast}}}}
\newcommand{\EELL}{\rm e^+ e^-\rightarrow l^{+}l^{-}}


\title{Putting non Point-like Behavior of Fundamental Particles to Test}

\author{ Irina Dymnikova$^*$, Alexander Sakharov${^\dagger}$,
         J\"urgen Ulbricht${^\dagger}$ and Jiawei Zhao$^{\ddagger}$}
\address{$^*$Institute of Mathematics and Informatics, UWM
in Olsztyn, PL--10-561 Olsztyn, Poland\\
$^{\dagger}$Labor f\"ur H\"ochenergiephysik, ETH-H\"onggerberg,
HPK--Geb\"aude, CH--8093 Z\"urich, Switzerland\\
$^{\ddagger}$Chinese University of Science and Technology, USTC,
Anhui 230029 Hefei, P.R.China}

\maketitle

\begin{abstract}

We review the experimental limits on those hypothetical interactions where the fundamental particles could
exhibit non point-like behavior. In  particular we have focused on the QED reaction
measuring the
differential cross sections for the process
$ \EEGG $
at energies around
91 GeV and 209 GeV with data collected from
the L3 detector from 1991 to 2001. With a global fit L3 set lower
limits at $ 95 \% $ CL on
a contact interaction energy scale parameter
$\Lambda > 1.6 $ TeV, which restricts the characteristic QED size of the
interaction region to $ R_{e} < 1.2 \times 10^{-17} $ cm.
All the interaction regions are found to be smaller than the Compton wavelength
of the fundamental particles. This constraint we use to estimate
a lower limit on the internal density of particle-like structure
with the de Sitter vacuum core. Some applications of obtained limits to the string and quantum gravity scales
are also discussed.
\end{abstract}

\section*{Introduction}
 When one starts to think about unification
of all known interactions the question whether the quarks and leptons are  structureless, or
 we will find that they have an extended structure, becomes very important. The point is that the
renormalization procedure
which allows to extract finite predictions for processes involving the strong,
weak and electro--magnetic interactions fails when gravitational interaction is
taken into account. Thus we are forced to use frameworks like the string theory ~\cite{stringtheory} to
incorporate gravity with other interactions.  As an essential part of string theory is that
the particles of standard model must have an extended structure. The strings
have new degrees of freedom that often take the classical geometry description
of propagation in extra dimensions. This means that, there will be additional
modifications of standard model amplitudes. In particular, some of the string
excitations could be visible in low energetic limit as a contact interaction
\cite{anton,string}. Some other phenomenological approaches, dealing with the explanation of mass spectrum of fermions
families (see for example \cite{Sakharov:pr} and references therein) or De Sitter-Schwarzschild behavior of particle--like objects
\cite{particle}, to which we pay particular attention in this paper, also could require non point--likeness of
fundamental particles\footnote{Some alternative, non gauge approach, to the explanation of the mass hierarchy of fermions families can be found in \cite{JUERGENSPG}.}. Thus it turns to be very important to study  possible experimental signatures of
operators causing the non point--like behavior of electron.

To test the size of fundamental particles (FP) two
experimental approaches have been developed. In
one approach a search is performed for excited
states of FP and corresponding mass is estimated \cite{LITKE01}.
This test would indicate a new substructure of the
FP. In the second approach a characteristic scale
parameter $ \Lambda $ \cite{EBOLI01}
or formfactor $ F $ \cite{BOURILKOV}
is determined, constraining the
characteristic size of the interaction region
for the reaction \cite{OSAKA}.
All these tests search for
physics beyond the Standard Model (SM).

In the first part of this paper
we summarize the stringent experimental restrictions
set on the mass of excidet states of FP
\cite{L3data2002,LEP2002,CDF,UA2_EXQ}. The obtained limits on the
parameters of exited states and contact interaction
set bounds on possible extensions of interactions areas
and hence on the sizes of FPs. It turns out that all these limits are much smaller
than the Compton wavelengths \( \lambda \!\!\!\!- _{c} \) of FPs.

Further we apply the experimental limits, obtained
for exited states of FP, to study TeV scale quantum gravity
in superstring theory framework, like it has been proposed in \cite{anton,string}.

The rest of the paper is dedicated to the modeling of an
extended particle-like objects by de
Sitter-Schwarzschild self-gravitating core \cite{IRINA1}.

\section*{EXPERIMENTAL LIMITS ON THE SIZES
     OF FUNDAMENTAL PARTICLES }

To test the finite size of fundamental particles,
experiments are performed to search for compositeness,
to investigate a non-pointlike
behavior or form factors in strong, electromagnetic and electroweak
interactions. Each interaction
is assumed to have its characteristic energy scale
related to the characteristic size of interaction region.
In the following sub-sections we review the experimental limits
on excited particle masses, energy scales $\Lambda$ and form factors
R for all three interactions.

\bigskip

{\bf STRONG INTERACTION-}
To test  the color charge of the quarks,
the entrance channel
and the exit channels of the reaction
in the scattering experiment should
be dominated by the strong interaction. This condition
is fulfilled by the
 CDF $ p\bar{p} $
data \cite{CDF} which exclude excited quarks $ q^{*} $ with
a mass between $80$ and $570$~GeV at 95\%~CL. The
UA2 data \cite{UA2_EXQ} exclude $u^{*}$ and $d^{*}$ quark
masses smaller than $ 288 $~GeV at 90\%~CL.
In this case characteristic energy scale is given by
the mass of the excited quark. Associated characteristic size is
$r_q \sim \hbar/(m_{q}^* c)<3.5\times 10^{-17}$ cm.

\bigskip

{\bf ELECTROMAGNETIC INTERACTION-}
In the case of electromagnetic interaction
the process $ \EEGG $ is ideal to test the QED because
it is not interfered by the $ Z^{o} $ decay \cite{QEDclean}.
This reaction proceeds via the exchange of a virtual electron
in the t - and u - channels, while the s - channel is
forbidden due to angular momentum conservation.
Total and differential cross-sections for the process $ \EEGG $,
are measured at the $\sqrt{s}$ energies from 91 GeV to 209 GeV
using the data collected with the L3 detector from 1991 to 2001
\cite{L3data2002}. Similar studies at $\sqrt{s}$ up to
202 GeV were reported by ALEPH, DELPHI and OPAL \cite{LEP2002}

To search for a deviations from the Standard Model,
the agreement between the measured cross section
and the QED predictions is used to constrain
the existence of an excited
electron of mass $ m_{e^{*}} $ which
replaces the virtual
electron in the QED process \cite{LITKE01},
or to constrain
a model with deviation from QED arising from an
effective interaction with non-standard
$ e^{+} e^{-} \gamma $ couplings and
$ e^{+} e^{-} \gamma \gamma $ contact terms \cite{EBOLI01}.

In the case of the heavy exited electron
the effective Lagrangian
\begin{equation}
{\cal L}_{\textrm{excited}}=\frac{e}{2\Lambda_{e^{*}}}
\overline{\psi_{e^{*}}}\sigma^{\mu\nu}(1 \pm \gamma^{5})\psi_{e}F^{\mu\nu}
+ h.c.
\label{L-exited-e}
\end{equation}
is used. Where
$ \Lambda_{e^{*}} $ is the effective scale of the interaction,
$ F^{\mu\nu} $ the electromagnetic field tensor,
$ \psi_{e^{*}} $ and $ \psi_{e} $ are the wave function
of the heavy electron and the electron respectively.

In the case of the
non-standard coupling
a cut-off parameter $ \Lambda $
is introduced to parameterize the scale of the interaction
with the following effective Lagrangian,

\begin{equation}
{\cal L}_{\textrm{contact}}=i\overline{\psi_{e}}
\gamma_{\mu}(D_{\nu}\psi_{e})
(\frac{\sqrt{4\pi}}{\Lambda^{2}_{6}}F^{\mu\nu}+
\frac{\sqrt{4\pi}}{\tilde{\Lambda}^{2}_{6}}\tilde{F}^{\mu\nu})
\label{L-contact-e}
\end{equation}
the Lagrangian chosen in this case has an operator
of dimension 6, the wave function of the electrons
is $ \psi_{e} $, the QED covariant derivative is $ D_{\nu} $,
the tilde on $ \tilde{\Lambda}_{6} $ and $ \tilde{F}^{\mu\nu} $
stands for dual.

Setting the interaction scale $ \Lambda_{e^{*}} $ to $ m_{e^{*}} $
L3 derived at 95\% CL a lower
limit for an exited electron mass
to $\MESTAR >0.31$~TeV \cite{L3data2002,Bajo:2000nh}.
Also L3 performed in the case of non-pointlike
coupling an overall fit of the last years data
\cite{L3data2002,L3data1996,L3data1997,L3data2000,Bajo:2000nh}.

A lower limit at 95\% CL for the cut-off
parameter $\Lambda$ limiting the scale
of the interaction region of $ \Lambda > 1.6 $ TeV
is reported \cite{L3data2002}.

Characteristic size related to the case of interaction via
excited heavy electron is
$r_e\sim{\hbar / (m_{e^*}c) < } 6 \times 10^{-17} $ cm.
In the case of direct contact term interaction
$r_e \sim{ {(\hbar c )}/{\Lambda}}=1.2\times 10^{-17}$cm.
The size of the interaction region must be smaller than $r_e$ because
the behavior of fit's as functions of $ \Lambda $
indicated a limit only.

\bigskip

{\bf ELECTROWEAK INTERACTION-}
The $ ep $ accelerator HERA and the $ e^{+}e^{-} $
accelerator LEP  searched for excited and
non-pointlike couplings of quarks and leptons.
In particular LEP tested the compositeness
of quarks and leptons with form factors.
In the entrance channel the reaction proceeds
via magnetic and weak interaction and in the exit channel
all three interaction participate. In the following we focus
separately on quarks and leptons cases.

{\bf Excited and non-pointlike quarks-}
The electron-proton interaction at high energies
allows us to search for excited quarks.
The magnetic transition coupling of quarks
includes  a single production of excited quarks
through $t-$channel gauge boson exchange between
the incoming electrons and quarks.
At the LEP, excited quarks could be
produced via a $ Z^{0}, \gamma $
coupling to fermions.
No signal was found in both cases
\cite{H1_EXQ,ZEUS_EXQ,ALEPH_EXQ,L3_EXQ,OPAL_EXQ}.

The L3  searched for new effects involving
four fermion vertices contact interactions in all exit channels at
center-of-mass energies between $ 130 $ GeV
and $ 172 $ GeV \cite{L3_FERMP}.
As in the case of the QED
contact interaction, an effective Lagrangian
is introduced \cite{BOUNDS_CON1}:

\begin{equation}
{\cal L}_{\text{contact}}=\frac{1}{1+\delta_{ef}}
\sum_{i,j=L,R} \eta_{ij}
\frac{g^{2}}{\Lambda^{2}_{ij}}
(\bar{e}_{i}\gamma^{\mu}e_{i})(\bar{f}_{j}\gamma^{\mu}f_{j})
\label{L-fermion-contact}
\end{equation}

The four-fermion contact interaction
is characterized by a coupling strength, g,
and by an energy scale  $ \Lambda $.
The Kronecker symbol $ \delta_{ef} $
is zero except for the $ e^{+} e^{-} $ final
state when it is equal to 1. The parameter
$ \eta_{ij} $ defines the contact interaction
model by choosing the helicity amplitudes which
contribute to the reaction
$ e^{+} e^{-}\rightarrow f \bar{f} $. The
wave function $ e_{i} $ and $ f_{j} $
denote the left- and right-handed
initial-state electron and final-state fermion.
The value of $ g/\Lambda $ determines the characteristic scale
of the expected effects. In a general search
the energy scale $ \Lambda $ is chosen by
convention such that $ g^{2}/4\pi = 1 $
and $ | \eta_{ij} |=1 $ or $ | \eta_{ij} |=0 $
is satisfied.

Four helicity amplitudes $ \eta_{LL} $,
$ \eta_{RR} $, $ \eta_{LR} $ and
$ \eta_{RL} $ are investigated for nine
different models each. Limits in the
$ f \bar{f} $ final state range from
$ \Lambda > 2.5 $ TeV to $ \Lambda > 7.1 $ TeV,
for $ q \bar{q} $ from
$ \Lambda > 2.0 $ TeV to $ \Lambda > 4.3 $ TeV,
for $ u \bar{u} $ from
$ \Lambda > 1.2 $ TeV to $ \Lambda > 4.3 $ TeV and
for $ d \bar{d} $ from
$ \Lambda > 1.4$ TeV to $ \Lambda > 3.5 $ TeV.
These scales allow us to estimate an upper limit
on characteristic size
$ r_{q} $
related to strong interaction
of the quarks. Depending on the different
helicity amplitudes and models this scale
ranges from
$ R_{q} < 1.6 \times 10^{-17} $ cm to
$ R_{q} < 2.8 \times 10^{-18} $ cm.

Probing the compositeness via the form factor operators
an another type of effective Lagrangian incorporating the
properties of non pointlike interaction can be constructed from the high
dimension operators with derivatives \cite{Hall}.
The simplest explicit form of this Lagrangian reads
\cite{Koepp}
\begin{equation}
{L_{eff}=\frac{1}{\Lambda^2}\left\{g_Lc_L\bar
f_L\gamma^{\mu}f\delta^{\nu}Z_{\mu\nu}+ g_Rc_R\bar
f_R\gamma^{\mu}f\delta^{\nu}Z_{\mu\nu}\right\} }
\label{L-formfactor}
\end{equation}
where $Z_{\mu\nu}$ are the $\gamma$, Z and W
fermion vertices, $\delta^{\nu}$ the derivatives, $g_{L,R}$ are the coupling
constants and $\Lambda$ the cut off parameter.
$c_L$  and $c_R$  can be incorporated in the cross section

\begin{equation}
{\frac{d\sigma (e^+e^-\rightarrow f\bar f)}{d\cos\theta}\propto
\left(\frac{d\sigma (e^+e^-\rightarrow f\bar
f)}{d\cos\theta}\right)_{SM}f(F_L,F_R)}
\label{Formfactor-cross}
\end{equation}
with the form factors

\begin{equation}
{F_{L,R}^{(e;f)}=(1+c_{L,R}^{(e;f)}\frac{Q^2}{\Lambda^2})}
\label{Formfactor}
\end{equation}
where $Q^2$ are the Mandelstam variables s or t for s- or
t-channel exchange.

The last upper limit on the fermion radii has been obtained from the L3
\cite{BOURILKOV,OSAKA}
data, assuming a single form factor for all fermions. In
particular for the $q\bar q$ final states. The upper limit
on the quark radius at 95\% confidence level is

\begin{equation}
{r_{q\bar q}=\frac{\sqrt{6}}{c \Lambda}\hbar <2.8\times 10^{-17}cm}
\label{qq-size}
\end{equation}

{\bf Excited and non-pointlike leptons-}
 \begin{figure} [b!] 
\center{
{\centering \begin{tabular}{cc}
\resizebox*{0.5\textwidth}{0.3\textheight}{\includegraphics{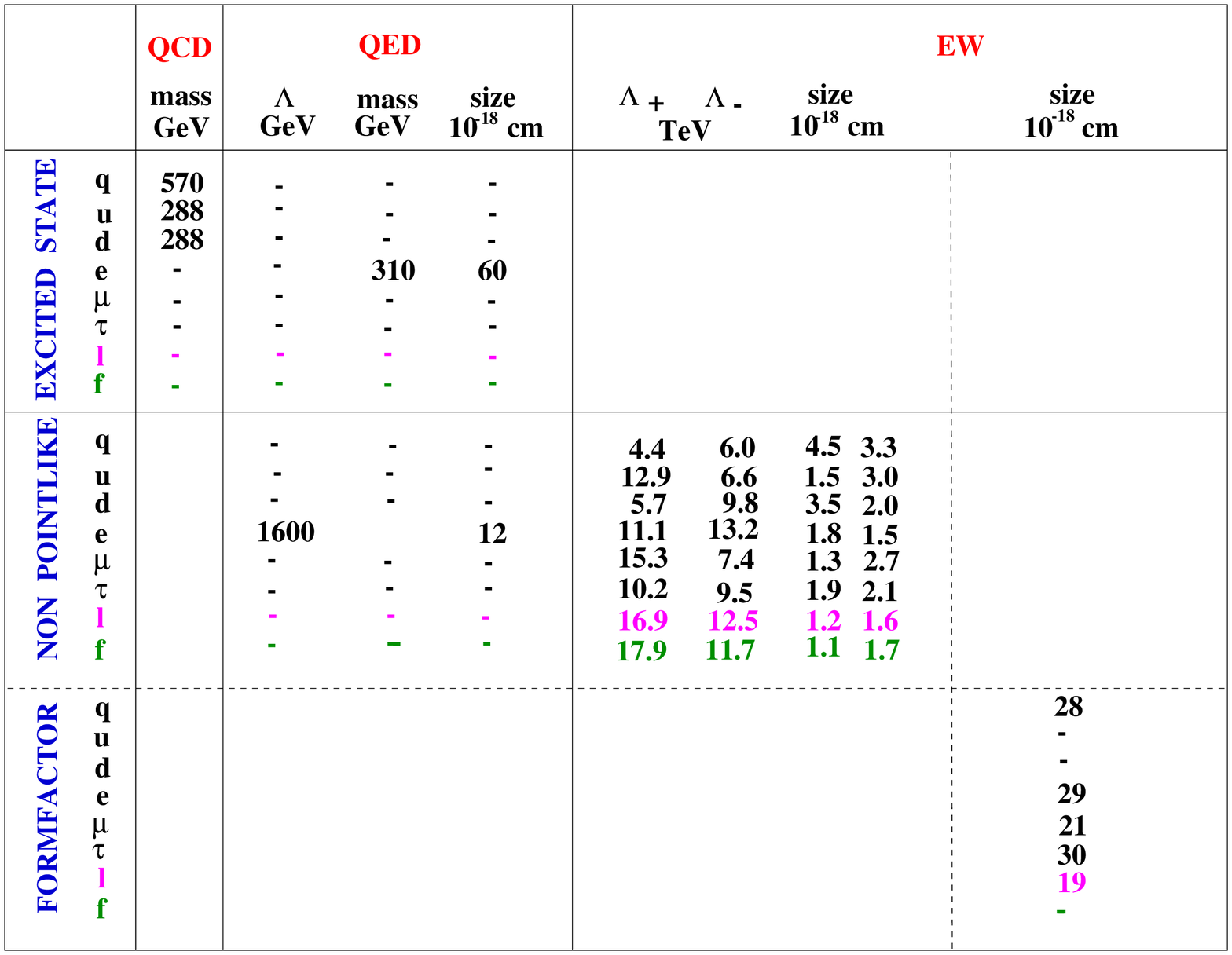}}&
\resizebox*{0.5\textwidth}{0.3\textheight}{\includegraphics{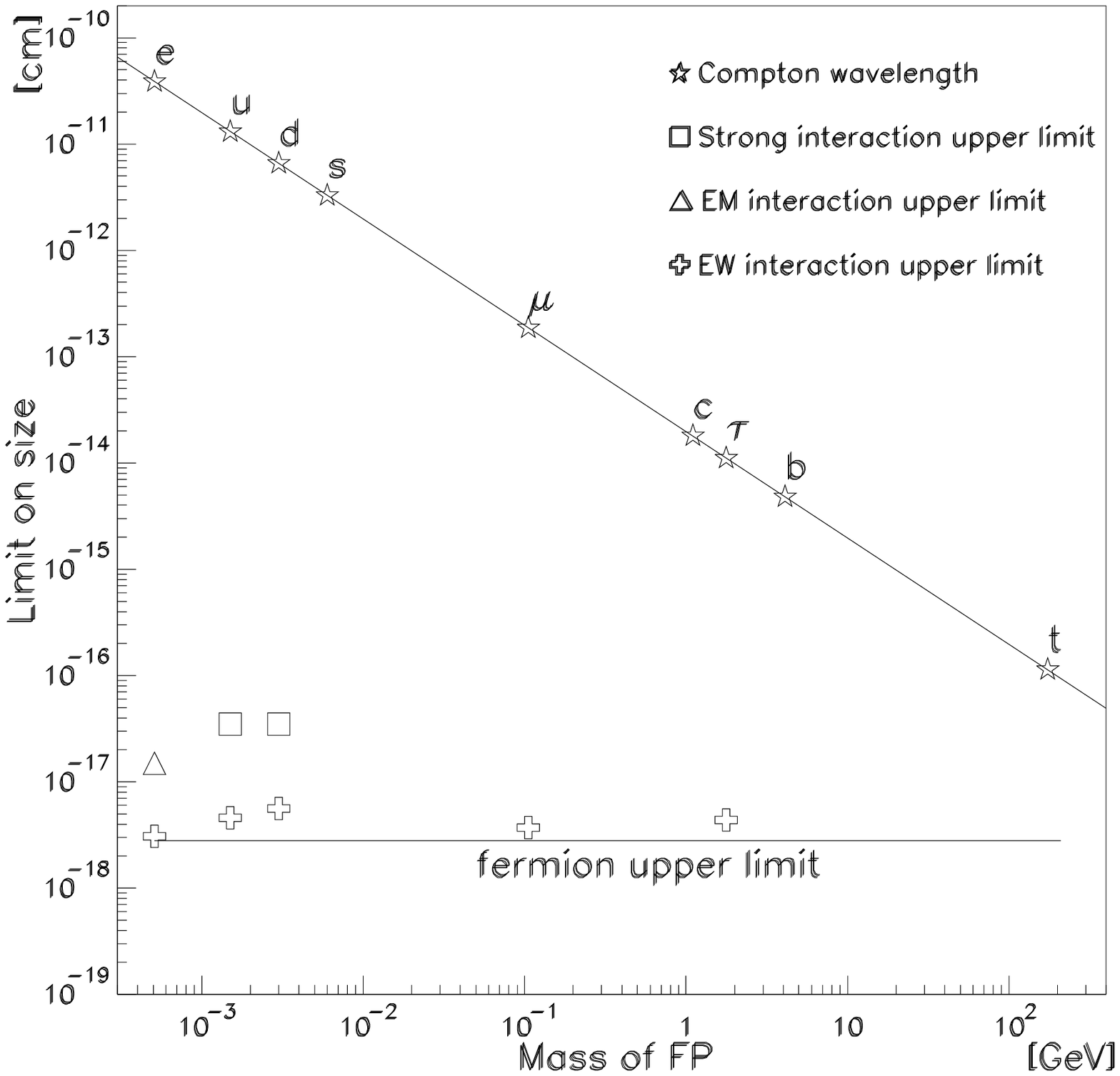}}\\
\end{tabular}\par}
}
\vspace{10pt}
\caption{In the left side the most stringent experimental
         limits of FPs are presented. The right side shows the
         comparison of Compton wavelengths of FPs with the
         current experimental limits measured for strong, electromagnetic
         and weak interaction.}
\label{summ}
\end{figure}
The electron-proton interaction at high energies
allows us to search, as in the quark case,
for excited leptons. For example excited electrons $ e^{*} $
and neutrinos $ \nu^{*} $ can be probed via the same
magnetic type coupling of the quarks.
At LEP, excited
leptons can be produced via
$ s- $ and $ t- $ channel for $ Z^{0}, \gamma $
and W
coupling to fermions, in particular to $ e^{*} $
$ \nu^{*} $.
No evidence for excited leptons was found \cite{H1_EXQ,ALEPH_EXQ,L3_2217}.

The L3 investigated the pure contact interaction
amplitudes $ e^{+} e^{-}\rightarrow l^{+} l^{-} $
and related deviations from the
Standard Model as in the quark case above \cite{L3_FERMP}.
Four helicity amplitudes $ \eta_{LL} $,
$ \eta_{RR} $, $ \eta_{LR} $ and
$ \eta_{RL} $ are investigated for nine
different models each. Limits in the
$ l^{+} l^{-} $ final state ranging from
$ \Lambda > 2.1 $ TeV to $ \Lambda > 7.1 $ TeV,
for $ e^{+} e^{-} $ from
$ \Lambda > 1.9 $ TeV to $ \Lambda > 6.4 $ TeV,
for $ \mu^{+} \mu^{-} $ from
$ \Lambda > 1.5 $ TeV to $ \Lambda > 5.3 $ TeV and
for $ \tau^{+} \tau^{-} $ from
$ \Lambda > 1.4$ TeV to $ \Lambda > 4.5 $ TeV.
These scales result in estimates of characteristic size
for weak interaction area $ R_{l} $
of the leptons. Depending on the different
helicity amplitudes and models this scale
ranges from
$ R_{l} < 1.4 \times 10^{-17} $ cm to
$ R_{l} < 2.8 \times 10^{-18} $ cm
as in the quark case.

The last upper limit on the lepton radii has been obtained from the L3
\cite{BOURILKOV,OSAKA}
data, assuming a single form factor for all fermions
in particular for the
$\mu^+\mu^-$ and $\tau^+\tau^-$ final states. The upper limit
on the electron radius at 95\% confidence level is
\begin{equation}
{r_{e}=\frac{\sqrt{6}}{c \Lambda}\hbar <2.9\times 10^{-17}cm}
\label{ee-size}
\end{equation}
All the experimental tests of the finite size of fundamental particles
so far have not shown any deviation from SM.
The experimental results
display in Fig. \ref{summ} on right side that all measured limits
of sizes of FPs are fare below the Compton wavelengths.
We will use this fact in the next chapter.

\section*{LIMITS TO THE STRING THEORY IN QUANTUM GRAVITY}
The  limits on the mass of excited electron obtained in \cite{L3data2002,Bajo:2000nh} is usefully to
study  TeV scale quantum gravity \cite{gia}. As it was recently suggested in
\cite{gia}, the fundamental scale of gravitational interaction $M$ cold be  as
low as TeV, whereas the observed weakness of the Newtonian coupling constant
$G_N\sim M_{Pl}^{-2}$ is due to the existence of N large ($\ell\gg\rm
TeV^{-1}$) extra dimensions into which the gravitational flux can spread out.
At the distances larger than the typical size of these extra dimensions the
gravity goes to its standard Einstein form, and  the usual Newtonian low can
be recovered via the relation $M_{Pl}=M^{N+2}\ell^N$ \cite{gia} between Plank
scale and scale $M$. It means that, such kind of quantum gravity becomes
strong at the energies $M$, where presumably all the interactions must unify,
without any hierarchy problem.

The phenomenological implications of large extra dimensions is
concentrated on the effects of real and virtual graviton emission. The
basic assumption is, that gravitons can propagate in extra dimensions \cite{gia}. The
quantum states of such gravitons are characterized by quantized momentum
in the large extra dimensions.

In the same time, the only known
framework that allows a selfconsistent description of quantum gravity
is string theory. As an essential part of the structure of string theory
\cite{stringtheory} is that the gravitons and the particles of standard
model must have an extended structure. This means that, there will be
additional modifications of standard model amplitudes due to string
excitations which can compete with or even overwhelm the modifications due
to graviton exchange \cite{anton,string}. An important effects of simple model
of string theory with large extra dimensions \cite{string} come from the
exchange of string Regge (SR) excitations of standard particles. In
standard model scattering processes, contact interactions due to SR
exchange produce their own characteristic effects in differential cross
section, and these effects typically dominate the effects due to Kaluza --
Klein (KK) exchange \cite{gia}. The SR excitation effects can be visible
as contact interactions \cite{anton,string} well below the string scale $M_S$.
The deviation from the standard model, we investigated, has been performed
in the terms of Drell's parameterization (ref.\cite{EBOLI01}). Actually, this
parameterization is applicable to any beyond standard model at short
distances. It was turned out, that from the comparison
of string theory result \cite{string}
to the (ref.\cite{EBOLI01}) the following identification can be deduced: $
\Lambda =\left(\frac{12}{\pi^2}\right)^{1/4}M_S $.
The last updated limit on $\Lambda >415$~GeV \cite{L3data2002,Bajo:2000nh}
corresponds to $M_S>386$~GeV at 95\% C.L. Using the
connection between string scale and quantum gravity scale $M$
\cite{string} we find $M>1188$~GeV.

\section*{PARTICLE-LIKE STRUCTURE RELATED TO GRAVITY }

In this section we will discuss  the modeling of an
extended particle-like objects by de
Sitter-Schwarzschild self-gravitating core \cite{IRINA1}.
This issue is actually inspired by the efforts to
bound only gravitationally objects which have masses
comparable with the masses of FPs.  Indeed, the size a
FP cannot be defined by
the Schwarzschild gravitational radius
$r_g=2G m c^{-2}$ ($m$ is the gravitational mass as measured
by a distant observer).
A size is constrained from below by the Planck length
$l_{Pl} \sim 10 ^ {-33}$ cm, and for any elementary particle its
Schwarzschild radius $r_g$
is many orders of magnitude smaller than $l_{Pl}$.
The Schwarzschild gravitational radius comes
from the Schwarzschild solution which implies point-like mass
and is singular at $r=0$.
The Schwarzschild metric can be modified by replacing a singularity
with  de Sitter regular core
\cite{IRINA1,werner,numerical,particle,lambda}.
This modified solution, de Sitter-Schwarzschild geometry, depends on the
limiting vacuum density $\rho_{vac}$ at $r=0$,
satisfying the equation of state
$p=-\rho_{vac}$.
The idea goes back to the mid-60s papers by Sakharov
who suggested that $p=-\rho$ can arise at superhigh densities
\cite{sakharov}, by Gliner who interpreted it as the vacuum equation
of state and suggested that it can be achieved as a result of
a gravitational collapse \cite{gliner}, and by Zeldovich who
connected $\rho_{vac}$ with gravitational interaction of virtual
particles \cite{zeldovich}.

In the context
of spontaneous symmetry breaking $\rho_{vac}$ is related to the potential of
a scalar field  in its symmetric (false vacuum) phase.
In the context of
Einstein-Yang-Mills-Higgs (EYMH)  self-gravitating non-Abelian structures
including black holes, $\rho_{vac}$ is related
to symmetry restoration in the origin. In a neutral branch of EYMH black hole
solutions a non-Abelian structure can be approximated as a sphere of a uniform
vacuum density $\rho_{vac}$
whose radius is the Compton wavelength of the massive non-Abelian field
(see \cite{maeda} and references therein).

The exact analytic solution describing de Sitter-Schwarzschild geometry,
was found in the Ref.\cite{IRINA1}.  It appeared that
the critical value of  the mass
$m_{cr}$ exists which selects two types of objects described by
de Sitter-Schwarzschild geometry:
a  neutral non-singular black hole
for $m\geq m_{cr}$, and for $m <m_{cr}$ a neutral
self-gravitating particle-like
structure with de Sitter vacuum core related to its
gravitational mass \cite{particle}. This
fact is generic for de Sitter-Schwarzschild geometry.
This geometry has two characteristic surfaces.
The surface of zero scalar curvature $r=r_s$ defines the gravitational size
$r_s$ of a particle-like structure.
Now a mass is not point-like but distributed, and most of it is within
$r_s$. The surface of zero gravity, $r=r_c<r_s$, defines a size
of an inner vacuum core. Both of them are at the characteristic scale
$\sim{(m/\rho_{vac})^{1/3}}$.

{\bf Selfgravitating particle-like structure with de Sitter vacuum core-} De Sitter-Schwarzschild geometry has been studied
as describing a black hole whose singularity is replaced
with de Sitter core of some fundamental scale
\cite{IRINA1,werner,numerical,particle}.

 Several solutions
have been obtained by directly matching de Sitter metric inside to Schwarzschild
metric outside of a junction surface \cite{numerical}.
Typical for matched solutions is that they have a jump at the
junction surface since the O'Brien-Synge junction condition $T^{\mu\nu}n_{\nu}=0$
is violated there \cite{werner}.
The exact
analytical solution
avoiding this problem
for a neutral spherically symmetric black hole
with a regular de Sitter interior was found in the Ref. \cite{IRINA1}.

The main steps to find this solution are to insert the spherically
symmetric metric
\begin{equation}
ds^{2}=e^{\nu }c^{2}dt^{2}-e^{\mu }dr^{2}-r^{2}
(d\theta^{2}+sin^{2}\theta d\phi ^{2})
\label{eq.7a}
\end{equation}
into the Einstein equations
\( R _{\mu \nu }-\frac{1}{2}Rg_{\mu \nu }=\frac{8\pi G}{c^{4}}T_{\mu \nu }
\)
which then take the form
\begin{equation}
\frac{-e^{\mu }}{r^{2}}+\frac{\mu ^{\prime }
e^{-\mu}}{r}+\frac{1}{r^{2}}=\frac{8\pi G}{c^{4}}T_{t}^{t}
\label{eq.7b}
\end{equation}
\begin{equation}
\frac{-e^{\mu }}{r^{2}}-\frac{\nu ^{\prime }
e^{-\mu}}{r}+\frac{1}{r^{2}}=\frac{8\pi G}{c^{4}}T_{r}^{r}
\label{eq.7c}
\end{equation}
\begin{equation}
\frac{1}{2}e^{-\mu }(\nu ^{\prime \prime }+\frac{\nu ^{\prime 2}}{2}+\frac
{\nu ^{\prime }-\mu ^{\prime }}{r}-\frac{\nu ^{\prime }\mu ^
{\prime}}{2})=\frac{8
\pi G}{c^{4}}T_{\theta}^{\theta}=\frac{8\pi G}{c^{4}}T_{\phi}^{\phi}
\label{eq.7d}
\end{equation}
To match smoothly the de Sitter metric inside to the Schwarzschild metric
outside, the boundary conditions are imposed on the stress-energy
tensor such that \( T_{\mu \nu }\rightarrow 0 \) as \( r\rightarrow \infty \)
and \( T_{\mu \nu }\rightarrow \rho_{vac}g_{\mu\nu} \) as
 \( r\rightarrow 0  \), with \( \rho _{vac} \) as de Sitter vacuum density
at \( r = 0 \). For both de Sitter and Schwarzschild metrics the
condition \( \mu = -\nu \) is valid, which defines the class of spherically
symmetric solutions with the algebraic structure of the stress-energy tensor
\( T_{\mu \nu } \) such that
\begin{equation}
T_{t}^{t}=T_{r}^{r}\,\, \textrm{and} \,\, T_{\theta }^{\theta }=T_{\phi }^{\phi }
\label{eq.7e}
\end{equation}
The stress-energy tensor of this structure describes a spherically symmetric
vacuum,
invariant under the boosts in the radial direction (Lorentz rotations in $(r,t)$ plane)
\cite{IRINA1}, and can be interpreted as $r-$dependent cosmological term
\cite{dynamics}.
It smoothly
 connects the de Sitter vacuum at the origin with the Minkowski
vacuum at infinity, and satisfies
the equation of state \cite{IRINA1,werner}
\label{alleqs8}
\begin{equation}
p_r=-\rho;~~p_{\perp}=p_r+{r\over 2} {dp_r\over{dr}}
\label{eq.8}
\end{equation}
where $p_r=-T_r^r$ is the radial pressure and $p_{\perp}=-T_{\theta}^{\theta}$
is the tangential pressure. In this class of solutions
the metric Eq.(\ref{eq.7a}) takes the  form
\label{alleqs9}
\begin{equation}
ds^2=\biggl(1-\frac{R_g(r)}{r}\biggr) dt^2 -
\biggl(1-\frac{R_g(r)}{r}\biggr)^{-1} dr^2 - r^2 d{\Omega}^2,
\label{eq.9}
\end{equation}
where
 $d{\Omega}^2$ is the metric on the unit two-sphere,
and
\label{alleqs10}
\begin{equation}
R_g(r)=\frac{2GM(r)}{c^2};~ ~ M(r)= \frac{4\pi}{c^2}\int_0^r{T^t_t(r)r^2 dr}
\label{eq.10}
\end{equation}
In the model of Ref. \cite{IRINA1}
the density profile $T^t_t(r)= \rho(r) c^2$ has been chosen as
\label{alleqs12}
\begin{equation}
\rho = \rho_{vac} e^{ -4\pi\rho_{vac} r^3/3 m }
\label{eq.12}
\end{equation}
which describes, in the semiclassical limit, vacuum polarization in the gravitational field
\cite{particle}.
Inserting Eq.(\ref{eq.12}) into Eq.(\ref{eq.10}) shows that \( R_g(r) \) takes
the form
\label{alleqs13}
\begin{equation}
 R_{g}(r)=r_{g}(1-e^{-4\pi \rho_{vac}r^{3}/3m})
=r_g(1-e^{-r^3/r_0^2r_g})
\label{eq.13}
\end{equation}
where
\label{alleqs14}
\begin{equation}
r_0^2 = \frac{3 c^2}{ 8 \pi G \rho_{vac}}
\label{eq.14}
\end{equation}
is the de Sitter horizon,
$ r_g = 2 G m / c^2 $ is the Schwarzschild horizon, and \( m \) is
the gravitational mass of an object.
In the limit of \( r <\!\!< (r_{0}^{2}r_{g})^{1/3} \)
Eq.(\ref{eq.13})
shows that \( R_g \rightarrow r^3 / r^2_0 \), and the metric
of Eq.(\ref{eq.9}) takes the de Sitter form with
\begin{equation}
g_{tt}=1- \frac{R_g(r)}{r}=1 - \frac{r^2}{r_0^2}
\label{eq.15}
\end{equation}
In the de Sitter geometry the horizon $r_0$ bounds the causally connected region.
An observer at $r=0$ cannot get information from the region beyond the surface $r=r_0$.

For $r\gg (r_0^2 r_g)^{1/3}$ the metric takes the Schwarzschild form
\label{alleqs16}
\begin{equation}
g_{tt}= 1 - \frac{R_g(r)}{r} = 1 - \frac{r_g}{r}
\label{eq.16}
\end{equation}
in agreement with boundary conditions.

The metric \( g_{tt}\left( r\right)  \) is shown
in Fig.{\ref{fig.5}. The fundamental difference from the Schwarzschild case
is that de Sitter-Schwarzschild black hole has two horizons, the black hole
horizon $r_{+}$ and the internal Cauchy horizon $r_{-}$.
%
\begin{figure}[htbp]
\vspace{-8.0mm}
\begin{center}
 \includegraphics[width=11.0cm,height=7.0cm]{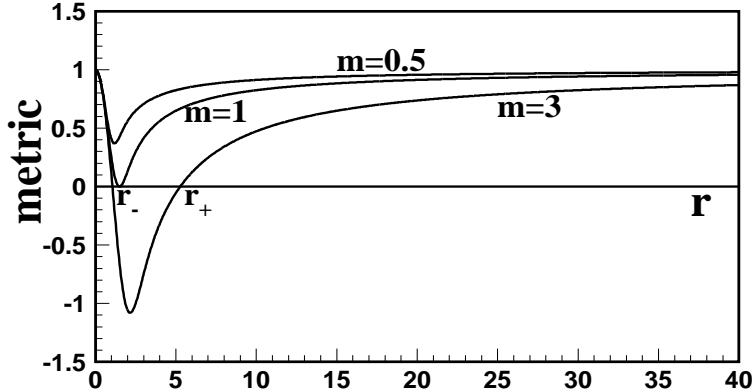}
\end{center}
\caption{De Sitter-Schwarzschild metric (\ref{eq.9})
         $g_{tt}(r)=1-R_g(r)/r$. The mass  $m$
         is  normalized to $m_{cr}$.
         The radius is  normalized to $r_{o}$.
         For $m>1$ we have a black hole solution,
         $m=1$ corresponds to the extreme
         black hole, and configuration with $m<1$ represents
         a recovered self-gravitating particle-like structure.}
\label{fig.5}
\end{figure}
The object is a black hole for
$m\geq m_{cr}\simeq{0.3 m_{Pl}\sqrt{\rho_{Pl}/\rho_{vac}}}$, which looses
energy via Hawking
radiation until a critical mass \( m_{cr} \) is reached where
the Hawking temperature
drops to zero \cite{particle}.
At this point the horizons come together. The critical value
$m_{cr}$
puts the lower limit for a black hole mass.
Below $m_{cr}$ de Sitter-Schwarzschild geometry Eq.(\ref{eq.9})
describes a neutral
self-gravitating particle-like structure made up of a
vacuum-like material Eq.(\ref{eq.7e})
with $T_{\mu\nu}\rightarrow \rho_{vac}g_{\mu\nu}$
at the origin \cite{particle}. This
fact does not depend on particular form of a density profile \cite{particle,dynamics}
which must only satisfy boundary conditions at the origin and at the infinity,
and guarantee
the finiteness of the  mass as measured by a distant observer
\label{alleqs17}
\begin{equation}
m = 4\pi \int_0^{\infty} {\rho(r) r^2 dr}
\label{eq.17}
\end{equation}
The interest of this paper is focused on the particle-like structure.
The case of $ m \geq m_{cr} $ is discussed in \cite{particle,dynamics}.

De Sitter-Schwarzschild geometry has two characteristic surfaces
at the characteristic
scale $ r \sim(r_{0}^2 r_{g})^{1/3}
 $\cite{particle}. The first is the surface of zero scalar curvature.
The scalar curvature
\( R = 8 \pi GT \) changes sign at the surface
\label{alleqs18}
\begin{equation}
r = r_s =
 \biggl ( \frac{m}{\pi \rho_{vac}}\biggr)^{1/3} =
 \frac{1}{\pi^{1/3}}   \biggl(\frac{m}{m_{Pl}}\biggr)^{1/3}
 \biggl(\frac{\rho_{Pl}}{\rho_{vac}}\biggr)^{1/3}l_{Pl}
\label{eq.18}
\end{equation}
which contains the most of the mass $m$.
Gravitational size  of a self-gravitating particle-like structure
can be defined by the radius $r_s$.
The second is related to
the strong energy condition of the
singularity theorems. It reads
$  (T_{\mu\nu} - g_{\mu\nu}T/2)u^{\mu}u^{\nu}\geq 0$,
where $ u^{\nu} $ is any time-like vector. The strong energy condition
is violated, i.e., gravitational acceleration changes sign, at the surface
of zero gravity
\label{alleqs19}
\begin{equation}
 r = r_c =
 \biggl ( \frac{m}{2\pi \rho_{vac}}\biggr)^{1/3} = \frac{1}
{(2\pi)^{1/3}} \biggl(\frac{m}{m_{Pl}}\biggr)^{1/3}
  \biggl(\frac{\rho_{Pl}}{\rho_{vac}}\biggr)^{1/3}l_{Pl}
\label{eq.19}
\end{equation}
The globally regular configuration with de Sitter
core instead of a singularity arises as a result
of the balance between attractive gravity
outside and repulsion inside of the
surface $r = r_c$.
This surface defines the characteristic
size of an inner vacuum core.
For a particle-like structure with $ m <\!\!< m_{Pl} $,
both these sizes are much bigger than
the Schwarzschild radius $r_g$. The ratio of a size of a  vacuum core
to the Schwarzschild radius $r_g$ is given by
\label{alleqs20}
\begin{equation}
\frac{r_c}{r_g}=\frac{1}{2}\frac{1}{(2\pi)^{1/3}}\biggl(\frac{m_{Pl}}{m}\biggr)^{2/3}
\biggl(\frac{\rho_{Pl}}{\rho_{vac}}\biggr)^{1/3}
\label{eq.20}
\end{equation}
 The horizons and characteristic surfaces
of de Sitter-Schwarzschild geometry are shown in the Fig.{\ref{fig.6}
where they are normalized to $r_0$.
\begin{figure}[htbp]
\vspace{-10.0mm}
\begin{center}
 \includegraphics[width=12.0cm,height=8.50cm]{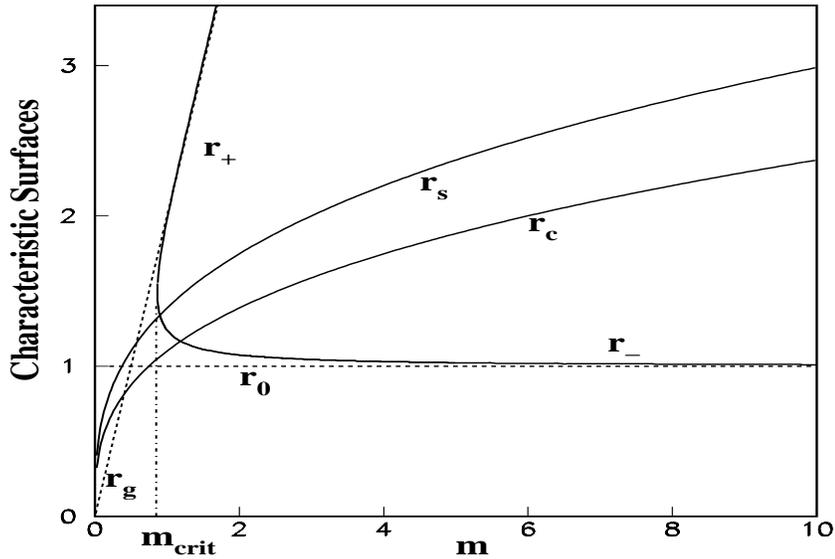}
\end{center}
\caption{Horizons $r_{\pm}$ and surfaces of zero curvature $r_s$ and zero
         gravity $r_c$ of de Sitter-Schwarzschild geometry. Schwarzschild
 radius $r_g$ and de Sitter radius $r_0$ are also shown.}
\label{fig.6}
\end{figure}

Thus one can see from Eq.(\ref{eq.18}), Eq.(\ref{eq.19}) and
Fig.\ref{fig.6}, that the proper matching of de Sitter and Schwarzschild geometry gives rise
for a stable particle-like structure. The size of such particle-like structure is defined by its mass \( m \)
and the vacuum density  $\rho_{vac}$ in the vicinity of \( r = 0 \).


{\bf Modeling of FPs with extended structure by an object with a vacuum interior-}
 Let us
consider a toy model of FP with extended structure represented by
a size of the De Sitter-Schwarzschild stable configuration described in the last paragraph.
It is natural to  believe that in the simplest realization of the De Sitter-Schwarzschild
particle-like object the energy density of a "vacuum-like
material" inside it can be attributed to the energy density of some scalar field. We
use Higgs ansatz to specify the potential of a scalar field governing by the vacuum interior


\begin{equation}
{V(\phi)}=-\frac{1}{2}\mu^{2}\phi^{2}+\frac{1}{4}\lambda\phi^{4}
           +\tilde{V}
\label{SCALAR}
\end{equation}
So the energy density inside the object is given by $\rho_{vac}=V(\phi )$ while
the term $\tilde{V}$ is added just to normalize the vacuum energy density
outside the object to the total density the of
the Universe, which is fixed by observations. 

The density profile of the vacuum core Fig.\ref{fig.5}. can be described by the following  function
\begin{equation}
 V(r)=V_{o} \cdot g_{tt}=V_{o}(1-R_{g}(r)/r)
\label{Self-Scalar1}
\end{equation}
where  $r$ is the distance from the center of
the object and $V_o=\rho_{vac}$ assigns the energy density of the scalar field in the
vicinity of the center of the vacuum core. Using the metric Eq.(\ref{eq.15}) deeply inside the vacuum interior one arrives to the following profile 


\begin{equation}
V(r \rightarrow 0)=
\rho_{vac}\cdot g_{tt}=\rho_{vac}\cdot(1-\frac{R_g(r)}{r})
= \rho_{vac}\cdot(1 - \frac{r^2}{r_0^2})
\label{eq.15a}
\end{equation}
Splitting the Eq.(\ref{eq.15a}) into three terms by the following manner
\begin{equation}
V(r \rightarrow 0)=
-\frac{1}{r r_{0}^{2}} \int_0^{r} {\rho(R) R^{2} dR } + \rho_{vac}
+C
\label{eq.15b}
\end{equation}
and taking into account the fact that  the integral in Eq.(\ref{eq.15b}) is just the mass integrated over the distance $ r $ from the center of the object.

\begin{equation}
 \int_0^{r} {\rho(R) R^{2} dR } = m(r)/4\pi ,
\label{eq.15c}
\end{equation}
we finally arrive to the following expression of the energy density in the vicinity of the center 

\begin{equation}
V(r \rightarrow 0)=
-\frac{1}{r r_{0}^{2}}\frac{1}{4\pi} m(r) + \rho_{vac}
+C.
\label{eq.15d}
\end{equation}
The last expression actually allows us to compare different contributions of the energy content of the vacuum interior with terms governing by the Higgs potential  Eq.(\ref{SCALAR}).


%


Let us do such comparison for the term $\lambda\phi^4$, which is responsible responsible for the selfinterraction of our scalar field. Taking as a pilot parameter the vacuum expectation value $v=246$~GeV measured in SM, one arrives to the following equation
\label{alleqs20a}
\begin{equation}
\rho _{vac}=\lambda v ^{4}/4
\label{eq.20a}
\end{equation}
From the other side  Eq.(\ref{eq.18}) if we believe in vacuum interior imposes the following relation  

\begin{equation}
 r_s =
 \biggl ( \frac{m}{\pi \rho_{vac}}\biggr)^{1/3},
\label{eq.18new}
\end{equation}
allowing in fact to calculate the coupling constant $\lambda$ if we know the size of the vacuum-like object we consider.
Recall, that in the experimental part of this paper we tested very carefully Fig.~\ref{summ}.  that
the gravitational size of FPs \( r_s \) confining most of its mass,
is restricted by its Compton
wavelength,

\begin{equation}
  r_s \leq \lambda \!\!\!\!-_c \ = \hbar /mc
\label{compton}
\end{equation}
This is also a natural assumption, since for a
quantum object \(\lambda \!\!\!\!-_c \)
constrains the region of its localization. Inserting Eq.(\ref{eq.18new}) and Eq.(\ref{eq.20a}) and 
into the condition Eq.(\ref{compton}) one can impose a limit
on self-interaction constant $ \lambda $ to
\label{alleqs21}
\begin{equation}
 1 \geq \frac{r_s}{\lambda \!\!\!\!-_c}=
\biggl(\frac{16\lambda}{\pi}\biggr)^{1/3}
\,  \, \textrm{;} \, \, \lambda \leq \frac{\pi }{16}
\label{eq.21}
\end{equation}
The limit Eq.(\ref{eq.21}) is also based on the assumption that the mass of thevacuum-like object we consider can be related to the parameters of
potential Eq.(\ref{SCALAR}) by the following manner
\begin{equation}
m=\sqrt{2\lambda v^{2}}= \sqrt{-2\mu^{2}},
\label{SCALAR mass}
\end{equation}
which is quite natural in our setup. Finaly we arrive to the conclusion that if we believe that an extended structure of FPs is constricted out of some a scalar field with the scale of symmetry breaking closed to the electroweak scale, the mass of the constituents of such vacuum interior (scalar field) can not exceed the level
\label{alleqs26}
\begin{equation}
m_{scalar}\leq 154 \, \textrm{GeV}
\label{eq.23a}
\end{equation}

In the framework
of our assumption the masses of  FPs are related to its gravitationally induced
core with de Sitter vacuum $\rho_{vac}$ at $r=0$.
This allows us to estimate the smallest size of  FPs as defined by
de Sitter-Schwarzschild geometry, a size of its vacuum core $r_c$, if we
know $\rho_{vac}$ and $m$. We assume that only one
mechanism exist in both models to generate the mass
of FPs, namely a particle gets its mass from the electroweak
scale $v$. Thus its inner core is determined by this scale.
Putting Eq.(\ref{eq.20a})
into the
Eq.(\ref{eq.19}) we get for a size of a vacuum core of a lepton with the mass $m_l$
\label{alleqs25}
\begin{equation}
{r_c}= \left(\frac{2m_l}{\pi\lambda v^4}\right)^{1/3}
\label{eq.22a}
\end{equation}
Then the constraint on $\lambda$ (Eq.(\ref{eq.21}))
 sets lower limits for the sizes of lepton vacuum cores by
$r_c^{(e)} > {1.5 \times {10^{-18}}}$ cm,
$r_c^{(\mu)} > {0.9 \times {10^{-17}}}$ cm, and
$r_c^{(\tau)} > {2.3 \times {10^{-17}}}$ cm.
Upper limits for sizes of vacuum cores  we estimate from the experimental
constraint on a Higgs mass $m_H > 107.0 $ GeV   \cite{higgs-up}. This gives
$r_c^{(e)} < 2.4 \times{10^{-18}}$cm,
$ r_c^{(\mu)} <  1.4 \times {10^{-17}}$cm, and
$ r_c^{(\tau)} < 3.6 \times {10^{-17}}$cm.


{\bf Most stringent limit to the sizes of leptons-} To make our consideration complete let us estimate the most stringent limit on $\rho_{vac}$
by taking into account that quantum region
of localization \( \lambda \!\!\!\!- _{c} \) must fit within a casually
connected region confined by the de Sitter horizon \( r_0 \). The requirement

\begin{equation}
   \lambda \!\!\!\!- _{c} \leq r_0
\label{limit}
\end{equation}
gives the limiting scale for a vacuum
density \( \rho _ {vac} \) related to a given mass \( m \)
\label{alleqs27}
\begin{equation}
\rho _ {vac} \leq \frac{3}{8 \pi} \left( \frac{m}{m_{Pl}}\right) ^{2}
\rho _ {Pl}
\label{eq.27}
\end{equation}
This condition connects a mass \( m \) with the scale for a vacuum density
\( \rho_{vac} \) at which this mass could be generated in principle, whichever
would be a mechanism for its generation.\\
\begin{figure}[htbp]
\vspace{-10.0mm}
\begin{center}
 \includegraphics[width=10.0cm,height=10.0cm]{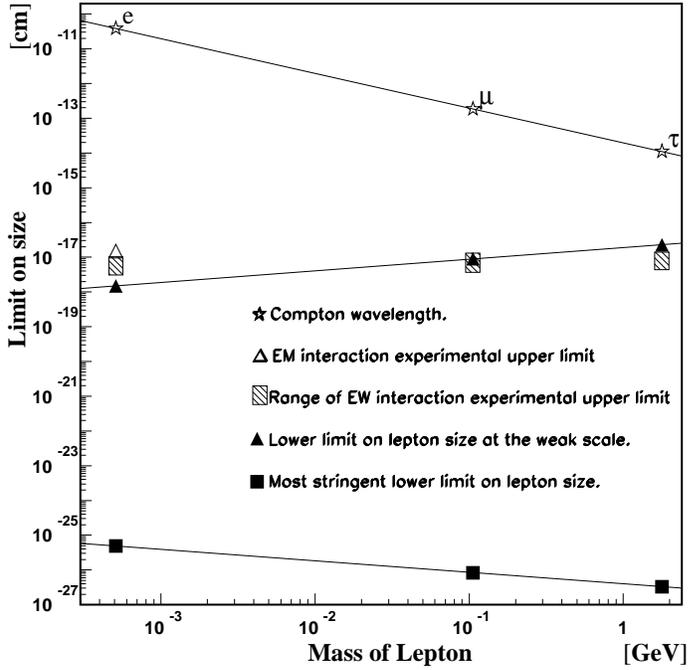}
\end{center}
\caption{Compton wavelength of leptons, experimental limits of
electromagnetic and weak interaction, and estimated lower limits for the size of leptons.
}
\label{fig.p3}
\end{figure}
In the case if FP have  inner
vacuum mass cores generated at the scale of Eq.(\ref{eq.27})
(for the electron this scale is of order of \( 4\times 10^{7} \) GeV),
we get from the Eq.(\ref{eq.27})
the most stringent, model-independent lower limit
for a size of vacuum core (Eq.(\ref{eq.19}))

\label{alleqs28}
\begin{equation}
r_{c}>\left( \frac{4}{3}\right) ^{1/3}\left( \frac{m_{Pl}}{m_l}\right)
^{1/3}l_{Pl}
\label{eq.28}
\end{equation}

Inserting the masses of the leptons $m_l$ into Eq.(\ref{eq.28}), we find
$r_c^{(e)} >
 4.9 \times 10^{-26} $ cm, $r_c^{(\mu)} >
 8.3 \times 10^{-27} $ cm, and $r_c^{(\tau)} >
 3.3 \times 10^{-27} $ cm.

The limits on the sizes of FP are summarized in Fig.\ref{fig.p3}, compared
to the Compton wavelength and to current experimental limits.
Let us emphasize that the most stringent limits
on sizes of FP as estimated in the frame
of de Sitter-Schwarzschild geometry, are much bigger
than the Planck length $l_{Pl}$. This fact supports our assumption
to compare the SM and de Sitter-Schwarzschild model what we
discussed at the beginning of this section.

Let us finally estimate an upper limit on a mass of a scalar at the energy
scale given by Eq.(\ref{eq.27}). For a scalar of ${\phi}^4$ theory
we put Eq.(\ref{eq.20a}) and $m_{scalar}=\sqrt{2\lambda}v$ into Eq.(\ref{eq.27})
and get the limit on the vacuum expectation value $v$, valid for any self-coupling $\lambda$
\begin{equation}
v\leq{\sqrt{\frac{3}{\pi}}m_{Pl}}
\label{eq.29}
\end{equation}
A self-coupling $\lambda$ for a scalar in ${\phi}^4$
theory is restricted by Eq.(\ref{eq.21}).
Then an upper limit for a scalar mass is

\begin{equation}
 m_{scalar}\leq{\sqrt{3/8}m_{Pl}}
\label{scalarplanck}
\end{equation}

These numbers give
constraints for the case of particle production
in the course of phase transitions in the very early universe. In this sense they give
the upper limits for relic scalar particles of ${\phi}^4$ theory.                \\

\section*{ CONCLUSION }

All the experimental tests of the finite size of fundamental particles
so far have not shown any deviation from SM. This is in particular the
case for excited states of fermions or non-pointlike behavior of fermions
in strong, electromagnetic and electroweak interactions.
This is demonstrated from the measured most stringent limits from excited states
of fermions, non-pointlike couplings and form factors.
The size of a FP can not only be defined by an interaction area $r$.
The wave character of the FP requests a characteristic wafelenght
of the FP. We test here the assumption that this
wafelenght is the Compton wavelengths \( \lambda\!\!\!\!-_c=\hbar/mc\).
All experiments confirm the assumption that the Compton wavelength
\( \lambda \!\!\!\!-_c \geq R \), the characteristic size of contact
interaction region of FP.

The experimental limits on the size of FP are in the case of strong
interaction for quarks $R_q <3.5\times 10^{-17}$ cm.
For the case of
pure QED interaction
the characteristic size for electrons $R_e$ is restricted by
$ \EEGG $ reaction
to $ R_{e} < 1.2 \times 10^{-17} $ cm.
The direct contact term measurements
for the electroweak interaction constrain
the characteristic size
for the quarks to $ R_{q} < 2.8 \times 10^{-18} $ cm,
and the leptons to
$ R_{l} < 2.8 \times 10^{-18} $ cm.
So far in all experiments no signal could be measured indicating a finite
size of FP.

The limits on the QED cut - off parameters $ \Lambda $ are used to
study TeV quantum gravity scale $ M $. The last L3
update shows that $ M > 1188 $ GeV.

All the actual limits depend only on the energy, luminosity of
the accelerator and the cross section of the reaction under investigation.

In the framework the modeling of FPs by de Sitter-Schwarzschild
 geometry wit vacuum interior governed by a Higgs scalar field the condition
\( \lambda \!\!\!\!-_c \geq R \)
restricts the self-coupling
of corresponding potential to
\( \lambda \leq \pi / 16 \). If the scale of
the generation of the FP masses is the electroweak scale
and if
we use the experimental limit for the Higgs \( m_H > 77.5 \) GeV
then our model constrains characteristic sizes of leptons to
$ 1.5\times {10^{-18}}$cm$< r_e < 2.4 \times{10^{-18}}$cm,
$ 0.9\times{10^{-17}}$cm$< r_{\mu}< 1.4 \times {10^{-17}}$cm, and
$ 2.3\times{ 10^{-17}}$cm$< r_{\tau} < 3.6 \times {10^{-17}}$cm.
The mass of the corresponding scalar should be under the level $m_{scalar} \leq 154 $ GeV.

Self-gravitating particle-like structure with de Sitter core
is generic. It is obtained from the Einstein equations with the boundary
conditions of the de Sitter vacuum at $r=0$ and Minkowski vacuum at
the infinity.
For the case of maximum possible scale for $\rho_{vac}$  at which a
particle could get its mass, it gives model independent  constraints
on sizes of vacuum cores for leptons which are
$r_e > 4.9 \times {10^{-26}}$ cm,
$r_{\mu} > 8.3 \times {10^{-27}}$ cm,
$r_{\tau} > 3.3 \times {10^{-27}}$ cm.
In the case of generation in the early de Sitter phase of the universe
\( m_{scalar} \leq \sqrt{3/8} m_{Pl} \).
The characteristic sizes as defined by  de Sitter-Schwarzschild
geometry are several order of magnitude bigger
than the Planck size, which justifies estimates for gravitational sizes
given in the frame of classical general relativity.

{\bf ACKNOWLEDGMENTS}

We are  grateful to Samuel C. C. Ting for his strong support of this project,
and to Martin Pohl for stimulating discussions of this paper.

\end{document}